\def\year{2021}\relax
\documentclass[letterpaper]{article} 
\usepackage{aaai21}  
\usepackage{times}  
\usepackage{helvet} 
\usepackage{courier}  
\usepackage[hyphens]{url}  
\usepackage{graphicx} 
\usepackage{csquotes}
\usepackage{rotating}
\usepackage{natbib}  
\usepackage{caption} 
\usepackage{array}

\usepackage{wrapfig}
\usepackage{multirow}
\usepackage{makecell}
\begin{document}
%


\title{Understanding Patterns of Users Who Repost Censored Posts on Weibo}

\author{Yichi Qian, Feng Yuan, Hanjia Lyu, Jiebo Luo\\}
\affiliations{
University of Rochester\\
\{yqian17, yfeng46, hlyu5\}ur.rochester.edu,jluo@cs.rochester.edu
}
\maketitle


\begin{abstract}
In this study, we focus on understanding patterns of  users whose repost contents would later be censored on weibo, a counterpart of Twitter in China as a social media platform. 
Little is known about the way regulations and censorship work in this indigenous platform and what role it plays in affecting users' expression of ideas. We collect over a million reposts from over 18,000 users and investigate the patterns of users whose reposts contain content that is no longer visible to the public, from the perspective of user location, device, gender, social capital as well as certified status. We find that user characteristics play an important role in affecting behaviors on Weibo. 
\end{abstract}

\section{Introduction}
The word ``JIA'' has become a prevailing verb on Chinese social media platforms, meaning that one's post has been censored or deleted against the composer's will. The appearance of this teasing verb signifies the frequency of posts being deleted or censored, and its consequential measures such as account being muted for a certain period of time or even the closure/suspension of the account. Usually, the censorship will take place if: 
\begin{enumerate}
    \item The post is related to pornographic content.
    \item The post is spreading misleading or false information.
    \item The post is about the subversion of government or calling on the protest or other social movements.
    \item The post is politically sensitive, i.e., mention of current leaders, corruption accusation on leadership, key social movements that have happened in history (Cultural revolution, Tiananmen square protest, and so on).
    \item The post is describing an ongoing protest.
    \item The post is sharing the news that the ruling government finds unsuitable to spread.
\end{enumerate}

As daily users of Weibo, we are deeply intrigued by such a phenomenon as we find that when we randomly open users' homepages, sometimes we could barely find the trace of such censorship while sometimes we might encounter a user whose full homepage is filled up with posts that repost censored contents. Therefore, we attempt to characterize the users whose repost contents contain the censored information. 
In particular, we focus on the following research question:
\begin{itemize}
    \item How does the pattern of reposting censored contents vary across user demographics?
\end{itemize}

To summarize, we characterize the Weibo users who repost censored posts, and examine the findings using a mixture of statistical methods. We find that users from economically developed areas are more likely to repost censored contents. Weibo users who have more social capital or certified Weibo users are less likely to share censored contents. We also find differences in the devices users use to make reposts. However, there is no gender bias with respect to sharing censored contents.

\section{Related Work}
The strict censorship on Chinese social media has long been a research focus ever since Twitter and Facebook were blocked in China in the late 2000s. Previous research on censorship deletion practices in Weibo has discussed the patterns of censorship as well as the comparison of deletion policies between Weibo and Twitter~\cite{bamman_o'connor_smith_2012}. Another study conducted by \citet{liu_zhao_2020} pointed out a novel insight that multimedia posts are more likely to experience censorship deletion than plain text posts. In our paper, we focus on characterizing the Weibo users who repost censored posts.

\section{Methods}
\subsection{Data Collection}
We used the ``weibo-crawler''\footnote{https://github.com/dataabc/weibo-crawler}, an open-source web crawler designed specifically for Weibo to collect the user characteristics of 18,231 Weibo users. The user characteristics are publicly available, including location, gender, whether this user is certified or not, and the number of followers. In addition, the publicly available reposts of each Weibo user are collected. The repost object we scrap from each user contains the original post, the repost with/without comments, and the device this user use to make reposts. After filtering out the accounts with no posts or no valid user information, 5,173 unique users are included in the final dataset.


\subsection{Preprocessing}
\subsubsection{Location.}
\begin{figure}[htbp!]
    \centering
    \includegraphics[width = 0.5\textwidth]{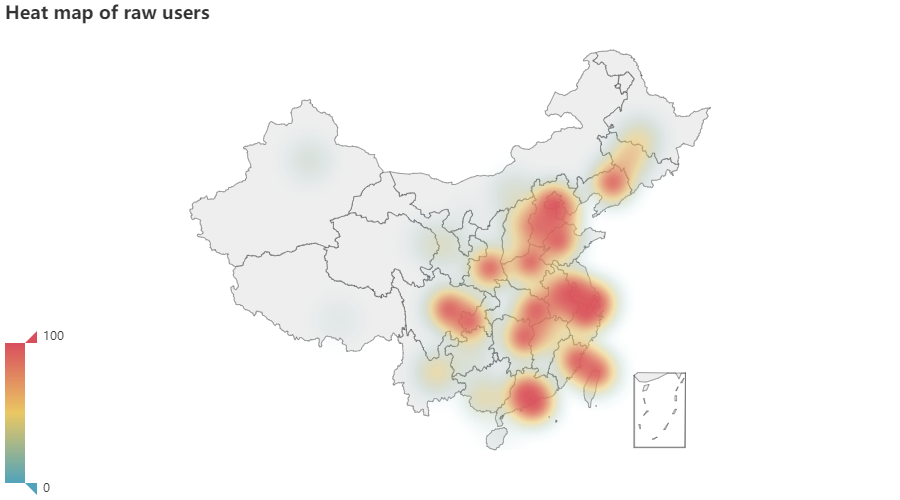}
    \caption{Heat map of the user distributions.}
    \label{fig:heat_raw}
\end{figure}

Figure~\ref{fig:heat_raw} shows the user location distribution of our dataset. The trend matches well with the natural population distribution and economic vitality in China, where Beijing, the Yangtze Delta, Guangdong-Hong Kong-Macao Greater Bay Area (GBA) and the middle-China economy belt area are the four economic centers with most Internet users as well. 

\subsubsection{Follower.}
The number of followers is transformed into a binary variable. If a Weibo user has more than 5,000 followers, the {\tt follower} variable will be labelled 1, otherwise it will be labelled 0. In our dataset, 81.5\% are labelled 1, and the rest are labelled 0. 
\subsubsection{Device.}
\begin{figure}[htbp!]
    \centering
    \includegraphics[width = 0.47\textwidth]{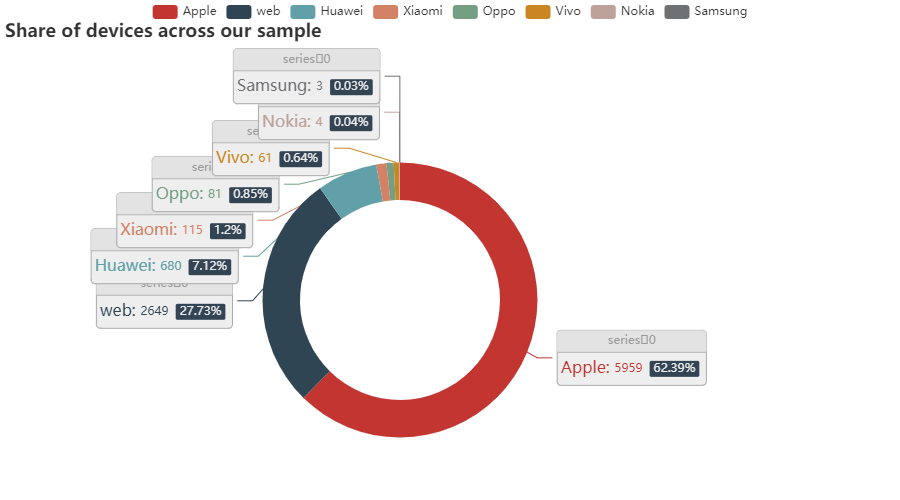}
    \caption{Share of devices.}
    \label{fig:raw_device}
\end{figure}

Users may have used multiple devices to make reposts, therefore we use the mode to represent the device of the user. All models of the same brand are merged. For example, {\tt honor}, {\tt nova} and {\tt mate} are merged into {\tt Huawei}. In the end, seven brands are included: {\tt Apple}, {\tt Huawei}, {\tt Xiaomi}, {\tt Oppo}, {\tt Vivo}, {\tt Samsung}, {\tt Nokia} and the rest are from a web browser (i.e., not from a phone). Figure~\ref{fig:raw_device} shows the share of device brands. {\tt Apple} is the most dominant brand on Weibo, with a share of 62.39\%, followed by web browsers, {\tt Huawei} (7.12\%) and {\tt Xiaomi} (1.20\%). All remaining brands combined constitute less than 1\%. The result is quite surprising because Huawei is a local brand and the sales data in the past quarters showed that Huawei has already taken the first place in terms of units sold worldwide~\cite{counterpoint_team}. The reason for such a huge discrepancy between the dominant device on Weibo and real life might be because that most heavy Internet users still pick {\tt Apple} as their top choice when buying cell phones. 

\subsubsection{Censored post.}
After examining a pilot batch of original posts, we find that if the post meets one of these three criteria,  this post has been censored or deleted: 1) It starts with the word ``Sorry''; 2) It contains the phrase ``This post has been reported for numerous times.''; 3) It contains the phrase ``This account has been reported for violating ...''. Therefore, we apply these three rules to the original posts of the reposts of each user and count the number of censored posts. In addition, we add a binary label to represent whether the user has ever reposted a censored post. In this way, the whole dataset is separated into two groups. For simplicity, we name the group of users who have ever reposted at least one censored posts {\tt Censored}, and the other {\tt Uncensored}.

\section{Results}
In this section, we intend to conduct a comparative analysis between {\tt Censored} and {\tt Uncensored} groups to characterize the Weibo users who repost censored posts.


\begin{figure}[htbp!]
\centering
\includegraphics[width=0.60\textwidth]{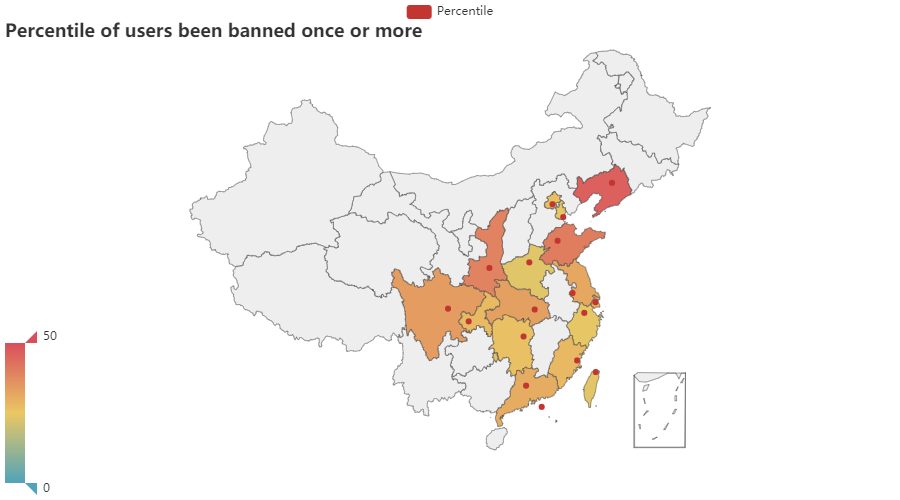}
    \caption{Percentage of users who repost posts that are later censored (by region).}
    \label{fig:percentile_map}
\end{figure}

\subsubsection{Users from economically developed areas are more likely to repost posts that are later censored.} Figure~\ref{fig:percentile_map} depicts the percentage of users who have reposted at least one post that is later censored. Provinces with fewer than 50 valid users are excluded. The trend mostly corresponds to our observation from Figure~\ref{fig:heat_raw}, but the provinces of Shanxi and Liaoning have shown a pattern that is contradicting to our previous presumption: the percentages of these two province are almost the same as those of Beijing and Shanghai. 
In the Greater Bay Area, Hong Kong is worth mentioning as it has the highest percentage among all major cities in China, with the percentage being over 33\%. It might be because that users from Hong Kong are more willing to discuss political issues and tend to repost news about their social movements or political ideas, which are sensitive to the Chinese authorities. The province of Hubei is also noteworthy: being the first place where the pandemic swept through, the people from Hubei were very expressive.

\begin{figure}[htbp!]
    \centering
    \includegraphics[width = 0.60\textwidth]{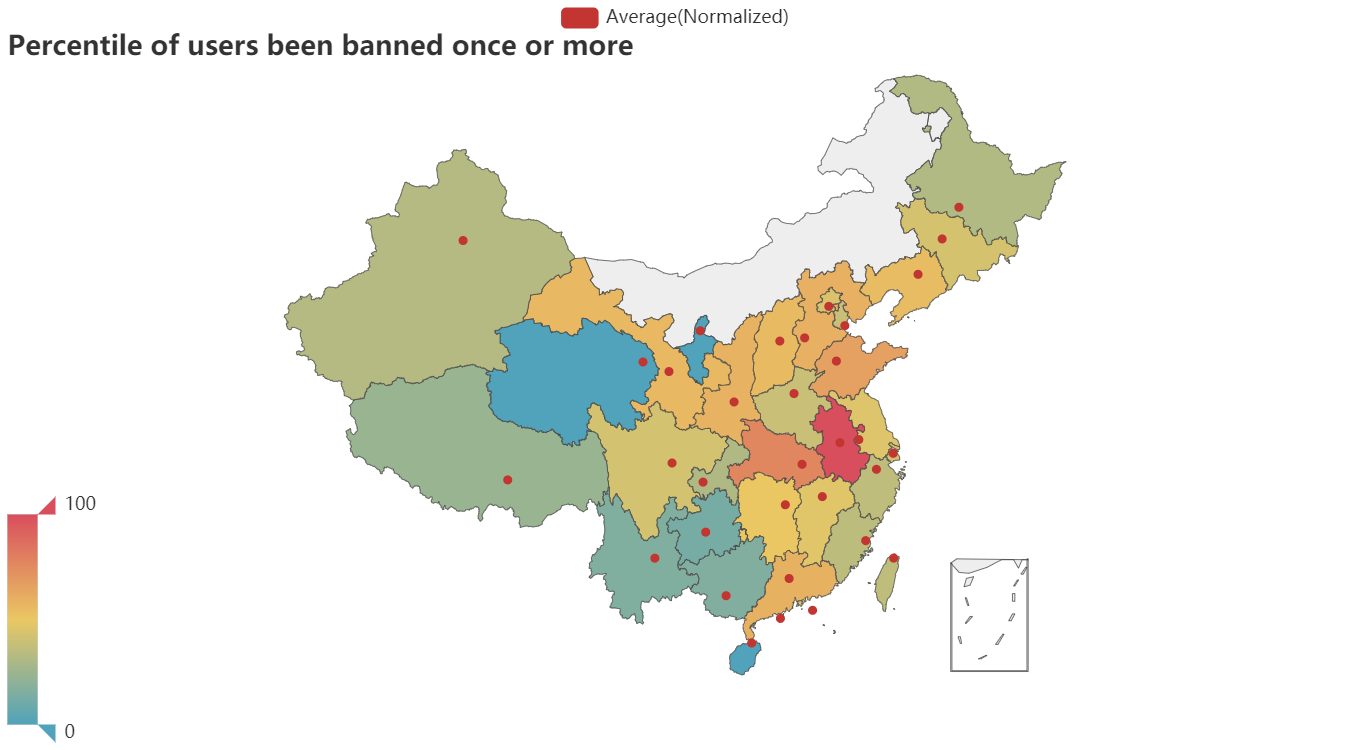}
    \caption{Normalized average censored reposts per user by region.}
    \label{fig:reposts_per_user}
\end{figure}

\subsubsection{Eastern and southern China regions have higher average censored reposts per user.} In Figure~\ref{fig:reposts_per_user}, we take the limit of valid users off, and plotted the average censored reposts per user for almost every province in China. Anhui province is the area where the average tops, which is much to our surprise. The reason  might be attributed to a small size of sample and some outliers that distorted the number. In general, it can be observed that a region is located in southern or eastern China, a high average censored reposts per user can be expected. One interesting phenomenon is that Hubei province has the second highest average in all provinces in mainland.


\subsubsection{There is no gender bias with respect to sharing censored contents.} 
\begin{figure}[htbp!]
    \centering
    \includegraphics[width = 0.47\textwidth]{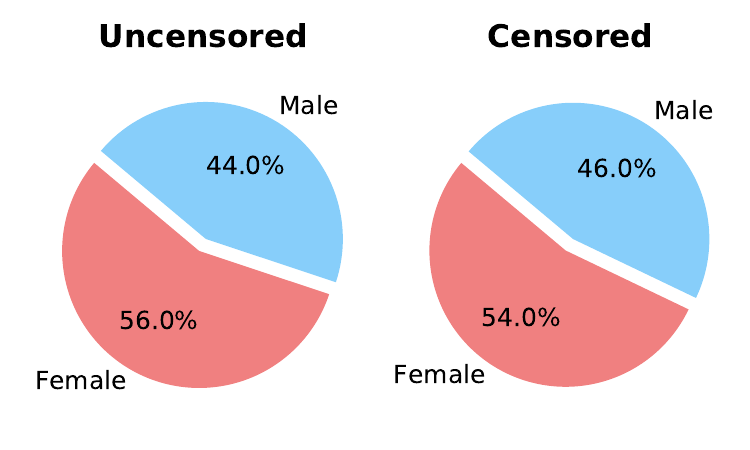}
    \caption{Gender distributions of {\tt Uncensored} / {\tt Censored} groups.}
    \label{fig:gender}
\end{figure}

Figure~\ref{fig:gender} depicts the gender distribution of {\tt Uncensored} / {\tt Censored} groups. From the pie chart we can tell that males seems to be the group which tends to repost contents that are censored later. 44\% of the users of the {\tt uncensored} group are male users and 46\% of the users of the {\tt censored} group are male users. However, after conducting the proportional z-test, we did not find significant differences in gender across two groups, indicating that there might not exist an actual difference. A case study of gender-based censorship on Facebook~\cite{nurik_2019} reveals that female users are more likely to experience inequality on social media platforms. In our case, the missing of such observation might be explained by the differences in cultures and censorship policies.

\subsubsection{Weibo users with more social capital are less likely to share censored contents.}
\begin{figure}[htbp!]
    \centering
    \includegraphics[width = 0.47\textwidth]{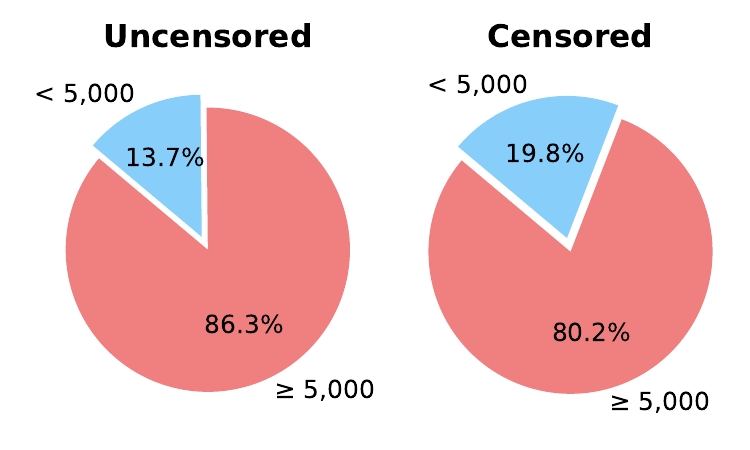}
    \caption{{\tt follower} distributions of the {\tt Uncensored} / {\tt Censored} groups.}
    \label{fig:fan_false}
\end{figure}

Figure~\ref{fig:fan_false} shows the {\tt follower} distributions of the {\tt Uncensored} / {\tt Censored} groups. 80.2\% of the users of {\tt Censored} have at least 5,000 followers, compared with 86.3\% of {\tt Uncensored}. After performing proportion z-test, we find there are proportionally more users who have at least 5,000 followers in {\tt Uncensored} ($p<0.05$), indicating that users with more social capital are less likely to repost censored contents.

\subsubsection{The pattern of reposting censored contents varies across devices.}
\begin{figure}[htbp!]
    \centering
    \includegraphics[width = 0.4\textwidth]{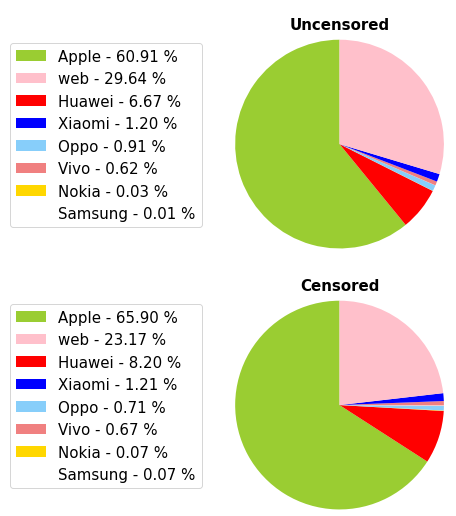}
    \caption{Device distributions of {\tt Uncensored} / {\tt Censored} groups.}
    \label{fig:brand_false}
\end{figure}

After performing the goodness-of-fit test, we find that the device distributions of {\tt Uncensored}/{\tt Censored} groups are significantly different ($p<0.05$) as shown in Figure~\ref{fig:brand_false}. The likelihood of users using \textit{Apple} to repost contents that are later censored is higher than the average. In terms of web browsers, users who mostly use web browsers are less likely to share contents that will be later censored. \textit{Huawei} users are more likely to repost censored contents than the average users but they do not repost as much such contents as the average users. The reason for such a scenario could be related to age and user occupations. \textit{Apple} has been extremely popular among students and people who work for the Internet industry. Web browsers, as relatively outdated tools, are mainly used by mid-aged users and they tend to be less involved in Internet hot-spot topics. For \textit{Huawei} users, many of them work in the public service sector so they tend to be mildly involved in the public-space discussion but they refrain themselves from further involvement. 

\subsubsection{Certified Weibo users are more cautious and share fewer censored contents.}
\begin{figure}[htbp!]
    \centering
    \includegraphics[width =0.47\textwidth]{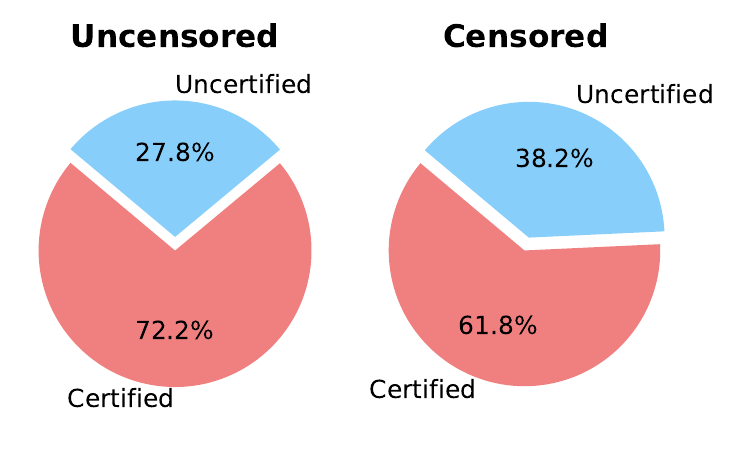}
    \caption{Distributions of the certified users of {\tt Uncensored} / {\tt Censored} groups.}
    \label{fig:certi_false}
\end{figure}

Figure~\ref{fig:certi_false} shows the proportions of the certified users of {\tt Uncensored} / {\tt Censored} groups. 61.8\% of the users of {\tt Censored} are certified, compared with 72.2\% of {\tt Uncensored}. After performing proportion z-test, we find sufficient evidence ($p<0.05$) to conclude that there are proportionally more certified users in {\tt Uncensored}, indicating that certified usersare less likely to repost censored contents.  An article written by \citet{fu_chan_chau_2013} suggest that the new real-name registration system enforced in China might have stopped some certified microbloggers from writing about social and political subjects, which is also consistent with our findings.

\section{Conclusion}

In this paper, we characterize Weibo users whose reposts contain contents that are later censored by the operator of Weibo. We find that users from economically vibrant areas have a higher tendency to repost contents that will be later censored. The finding that Hong Kong is the region with the highest percentage of users who have reposted contents that were later censored is no surprise due to its higher degree of freedom of speech and its recent social movements. We also find that users with more influential powers tend to be more cautious and prudent in reposting controversial contents. From the perspective of gender, male users are no more likely to repost contents that are later deleted than female users. In terms of the devices, {\tt Apple} seem to be the most dominant device brand for users to access Weibo. Users using {\tt Apple} have a higher likelihood to repost such contents and higher frequencies as well. 

\subsubsection{Limitations.} Our study has limitations. The first is that although we have collected over 600,000 posts from over 18,000 users, the cohort may still not be representative enough since there are over tens million Weibo users. In addition, the collection process of users are not completely random due to the fact that we could not fetch randomly generated IDs. The way we capture new user IDs is by randomly picking users that are followed by the previous account(s) we scrap. Such a crawling method could have a hidden bias in selecting a group of users to represent the whole user population. 
Certified users or users with more followers have a larger chance of being picked. Finally, although the contents of the original posts are not publicly available, the comments are. We could conduct content analyses of the comments to indirectly examine the thematic characteristics of the censored posts.

\bibliography{ref}

\end{document}